# Three Wave Hypothesis, Gear Model and the Rest Mass

M. I. Sanduk

*School of Engineering, Faculty of Engineering and Physical Science, CORA, University of Surrey, Guildford Surrey GU2 7XH, UK*
m.sanduk@surrey.ac.uk

Abstract: Three Wave Hypothesis (TWH) is of a relativistic quantum foundation. The formulation of TWH has perfect similarities with bevel gear model.
The rest mass is considered within TWH and as a consequence of similarity, between TWH and the gear model, it is found that rest mass frequency is related the frequency of touch point of the gear in relative to the large wheel. In regarding this frequency instead of the small wheel frequency, the gear model leads to four-vector representation of a particle.

*Keywords: Three wave hypothesis, Wave function, four vector, Rest mass, Gear, De Broglie wave.*

*PACS: 03.65, 03.75, 04.80*

# 1. Introduction

Since the early days of quantum mechanics, and after de Broglie's wave hypothesis, many physicists (in addition to de Broglie) tried to consider particle within the frame of wave phenomena. De Broglie Wave has been investigated experimentally whereas the theories of wave origin of particle were facing hard obstructions. In the second half of the last century many researchers have been considered two or three waves representations [1, 2, 3, 4, 5]. In addition to de Broglie wave there is Compton wave; others postulated a third wave. One of those hypothetical works is Three Wave Hypothesis (TWH) of Horodecki [5, 6]. It is a relativistic-quantum model; and relates massive particle to wave phenomena. The hypothesis assumed a dual wave in addition to Compton and de Broglie waves. So far there is no experimental evidence confirming TWH.

However, The TWH shows that:

1- There are two dispersive, waves (de Broglie and dual), and transformed Compton wave.

2- The recognized relativistic quantities by the lab observer ($\omega_C, a_C, \omega_{C_\circ}, a_{C_\circ}$ the angular frequency and radius, for transformed and rest stet related to Compton wave) are related to the product of dispersive parameters.

3- There are two relativistic representations [5, 6]. The first is in relative to de Broglie wave:

$$\omega_C = \pm \left( \omega_B^{\,2} + \omega_{C_\circ}^{\,2} \right)^{\frac{1}{2}} \quad (1\text{-}1),$$

and the other, to dual wave:

$$\omega_C = \pm \left( \omega_D^{\,2} - \omega_{C_\circ}^{\,2} \beta^{-2} \right)^{\frac{1}{2}} \quad (1\text{-}2).$$

The single angular frequency formula of TWH is [7]:

$$\left( \omega_D^{\,2} - \omega_B^{\,2} \right)^{\frac{1}{2}} = \pm (\omega_D \omega_B)^{\frac{1}{2}} \left( \mu_T - \frac{1}{\mu_T} \right)^{\frac{1}{2}} \quad (1\text{-}3).$$

Where $\omega_B, \omega_D, \omega_C$, are de Broglie frequency, dual wave frequency, and the transformed Compton frequency respectively. The same form may be obtained for radius representation.

The ratio of wave parameters is:

$$\frac{a_B}{a_D} = \frac{\omega_D}{\omega_B} = \mu_T = \left(\frac{1}{\beta}\right)^2 \quad (1\text{-}4)$$

Where $a_B, a_D, \beta$ are de Broglie wavelength, dual wavelength, and the ratio of velocity of particle to the velocity of light ($\beta = \frac{v}{c}$) respectively. The wave velocities are:

$$\omega_D a_D = \omega_B a_B = c \quad (1\text{-}5)$$

A work in 2007 showed a perfect similarity between TWH formulation and a gear model [7]. This mechanical model is virtual gear model (VGM).

However, TWH is of relativistic and quantum base, and is a combination of them. That combination (system) has perfect similarity with gear model. The present work and with aid of the similarity of formulation, tries to find a relationship between VGM and special relativity.

### 1.1. Gear Model

A previous work [7] exhibited perfect similarities between the system of equations of the TWH and a classical gear of two perpendicular wheels (as that is shown Fig.1-1). This is a virtual gear model (VGM), and it is a pure mathematical model. It has classical feature but it is not classical materialistic model.

The gear model is characterized by [7]:

1- The total angular frequency of the perpendicular wheels:

$$\left(\omega_1^2 + \omega_2^2\right)^{\frac{1}{2}} = \pm\omega_R \quad (1\text{-}6).$$

Similar formula for the radius ($a_R$) can be obtained.

2- The gear ratio:

$$\frac{a_2}{a_1} = \frac{\omega_1}{\omega_2} = \mu \quad (1\text{-}7).$$

3- The velocities:

$$\omega_R a_R = v_R \quad (1\text{-}8)$$

$$\omega_1 a_1 = \omega_2 a_2 = v_{1,2} \quad (1\text{-}9),$$

and

$$\omega_R a_1 = v_{1R} \text{ and } \omega_R a_2 = v_{2R} \quad (1\text{-}10)$$

where $\omega_1, \omega_2, \omega_R, a_1, a_2, v_R$ and $\mu$ are the first, second and the total angular frequencies, the first, second, radii of the wheels, linear velocity, then the gear ratio respectively. The limit of $\mu$ is $1 \leq \mu < \infty$. Representing gear model in terms of wheel frequencies only (without the total frequency), one may obtain [7]:

$$\left(\omega_1^2 - \omega_2^2\right)^{\frac{1}{2}} = \pm \left(\omega_1 \omega_2\right)^{\frac{1}{2}} \left(\mu - \frac{1}{\mu}\right)^{\frac{1}{2}} \quad (1\text{-}11).$$

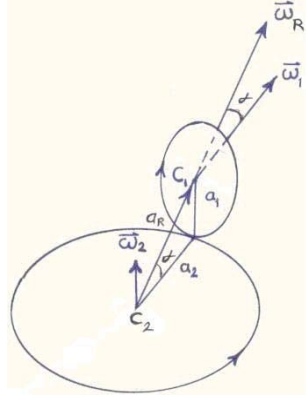

Fig. (1-1) The gear.

Same form may be obtained for radius representation.

**1.2. The similarity**

The similarity between the two models TWH and VGM is quite obvious in:
1- Ratio similarity, Eqs. (1-4) and (1-7).
2- Velocities similarity, Eqs. (1-5) and (1-9).
3- Single system representation, Eqs. (1-3) and (1-11).

*1.2.1 The correspondences*

The similarity may lead to the following correspondences:

1-
$$\omega_C \equiv (\omega_1 \omega_2)^{\frac{1}{2}} = \omega \quad (1\text{-}12).$$

Then, consequently (since $\omega_C^2 = \omega_D \omega_B$) one can say:

$\omega_B \equiv \omega_2$, $\omega_D \equiv \omega_1$, $a_B \equiv a_2$ and $a_D \equiv a_1$ (1-13).

2-
$$\omega_C \equiv \omega_R \sqrt{\frac{\mu}{1+\mu^2}} \text{ and } a_C \equiv a_R \sqrt{\frac{\mu}{1+\mu^2}} \quad (1\text{-}14),$$

where, $\omega_R$ and $a_R$ are the total angular frequency and the radii.

3-
$$c = a_C \omega_C \equiv v_R \left[\frac{\mu}{1+\mu^2}\right] \quad (1\text{-}15),$$

then:
$$v_D \equiv v_{1R} \sqrt{\frac{\mu}{1+\mu^2}} \equiv v, \text{ and } v_B \equiv v_{2R} \sqrt{\frac{\mu}{1+\mu^2}} \quad (1\text{-}16),$$

where $v_D, v_B, v_{1R}, v_{2R}, v$ are phase velocity of dual wave, phase velocity of de Broglie wave, velocity of first wheel, velocity of second wheel and the particle velocity respectively. It is easy to find that:

$$v_D v_B \equiv v_{1R} v_{2R} \left[\frac{\mu}{1+\mu^2}\right] \equiv c^2 \,(1\text{-}17).$$

The space of this VGM is three dimensional Euclidian space.

Since TWH is of quantum and relativistic base and this model exhibit similarity with VGM, so the gear model may lead to give explanations for both special relativity and quantum mechanics. Using VGM, the present work, tries to go in reverse direction to obtain the special relativity and quantum foundations.

## 2. The rest Compton frequency

According to TWH, Compton relative frequency is:

$$\omega_C = (\omega_D \omega_B)^{\frac{1}{2}} \quad (2\text{-}1)$$

It is related to a system of two waves. The rest Compton frequency, Eqs. (1-1) and (1-2), one can find:

$$[(\omega_D - \omega_B)\omega_B]^{\frac{1}{2}} = \omega_{C\circ} \qquad (2\text{-}2)$$

This is related to a system of two dispersive waves as well; but one of them $(\omega_D - \omega_B)$ is relative to the other.

Using the gear model similarity shows that:

$$(\omega_1 - \omega_2)\omega_2 = \omega_{xC_2}^{2} \qquad (2\text{-}3)$$

The quantity $\omega_{xC_2}$ has no analogy in gear system. The $(\omega_1 - \omega_2)$ is represented the frequency of first wheel relative to the second wheel, or:

$$\omega_1 - \omega_2 = \omega_{1C_2} \qquad (2\text{-}4)$$

$\omega_{1C_2}$ may represent the frequency of the touch point of the two wheels. Eq. (2-3) becomes:

$$\omega_{1C_2}\omega_2 = \omega_{xC_2}^{2} \qquad (2\text{-}5)$$

So, one can say that:

$$\omega_{C\circ} \equiv \omega_{xC_2} \qquad (2\text{-}6)$$

## 2.1 The relative gear

For lab observer the recognized quantities (with correspondent's fundamental gear) are (Eq.(1-1)):

$\omega_C = \omega \equiv (\omega_1 \omega_2)^{\frac{1}{2}}$, Compton frequency, and

$\omega_B = \dfrac{2\pi c}{\lambda_B} \equiv \omega_2$ de Broglie frequency.

Whereas $\omega_D \equiv \omega_1$, has no experimental evidence. So let, us try to reformulat the system without $\omega_1$. The new system is in term of $\omega$, $\omega_2$ and any other observable quantity.

In addition to Eq. (1-11) one may get many different forms for the relationship between $\omega_1$ and $\omega_2$. Using Eqs. (2-4) and (1-12), on gets:

$$\omega^2 = \omega_2^{2} + \omega_{xC_2}^{2} \qquad (2\text{-}6),$$

Eq.(2-6) represents a new gear or relative gear.

The new gear depicted in Fig.2-1, and its gear ratio (of Eq.(2-6)) is:

$$\frac{\omega_{xC_2}}{\omega_2} = \frac{a_2}{a_{xC_2}} = \mu_{C_2} \quad (2\text{-}10).$$

Where $\omega_{xC_2}, \omega_2, \omega, a_{xC_2}, a_2$ and $\mu_{C_2}$ are the first, second and the total angular frequencies, the first, second, radii of the wheels, then the gear ratio, respectively.

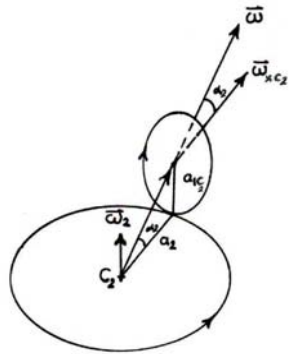

Fig (2-1) Relative gear

The parameters $\omega_{xC_2}, a_{xC_2}$ are functions of the independent gear wheels parameters $(\omega_1, \omega_2, a_1, a_2)$. The new gear ratio (Eq. (2-10)) $\mu_{C_2}$ in terms of the independent gear ratio ($\mu$) is:

$$\frac{\omega_{xC_2}}{\omega_2} = \frac{a_2}{a_{xC_2}} = \mu_{C_2} = \sqrt{\mu - 1} \quad (2\text{-}11).$$

From Eqs. (2-1) and (2-2) one obtains:

$$\omega_{xC_2} = \omega \left( \frac{1}{\frac{1}{\mu_{C_2}^2} + 1} \right)^{\frac{1}{2}} \quad (2\text{-}12)$$

or

$$\omega_{xC_2} = \omega \left( \frac{\mu - 1}{\mu} \right)^{\frac{1}{2}} \quad (2\text{-}13)$$

Owing the consideration of $\mu$ ratio of fundamental gear (Eq.(1-7)), and a relative consideration as well, so $\mu$ can be represented as:

$$\mu = \frac{a_2}{a_1} = \frac{v_{2C_2}}{v_{1C_2}} \quad (2\text{-}14),$$

where:

$$v_{1C_2} = a_1 \omega, \text{ and } v_{2C_2} = a_2 \omega \quad (2\text{-}15).$$

Eq. (2-13) can be rewritten as:

$$\frac{\omega}{\omega_{xC_2}} = \frac{1}{\sqrt{1 - \frac{v_{1C_2}}{v_{2C_2}}}} = \frac{1}{\sqrt{1 - \frac{v_{1C_2}^2}{v_{2C_2} v_{1C_2}}}} \quad (2\text{-}16),$$

and, for the radii:

$$\frac{a}{a_{xC_2}} = \sqrt{1 - \frac{v_{1C_2}^2}{v_{2C_2} v_{1C_2}}} \quad (2\text{-}17).$$

Let:

$$v_{2C_2} v_{1C_2} = \Gamma^2 \quad (2\text{-}18).$$

Then

$$\frac{\omega}{\omega_{xC_2}} = \frac{1}{\sqrt{1 - \frac{v_{1C_2}^2}{\Gamma^2}}} \quad (2\text{-}19)$$

$$\frac{a}{a_{xC_2}} = \sqrt{1 - \frac{v_{1C_2}^2}{\Gamma^2}} \quad (2\text{-}20)$$

Where $v_{1C_2}, v_{2C_2}, \Gamma$ are relative velocity of first wheel, second wheel and constant quantity (independent of the velocities) respectively. From Eqs. (2-17), (2-19) and (2-20), it is easy to find that:

$$\omega^2 \Gamma^2 - \omega^2 v_{1C_2}^2 = \omega_{xC_2}^2 \Gamma^2 \quad (2\text{-}21A),$$

and for radius:

$$a_{xC_2}^2 \Gamma^2 - a_{xC_2}^2 v_{1C_2}^2 = a^2 \Gamma^2 \quad (2\text{-}21B).$$

These forms are invariant forms (dot product of two 4-vectors), and in this case the $\omega_{xC_2}$ and $a_{xC_2}$ look as relativistic invariant quantities. The main reason behind this result is considering the system without $\omega_1$ and $a_1$.

**2.2. Similarity and correspondences**

In compression Eqs. (2-21) with special relativity representations, one finds:

$$\omega_{xC_2} \equiv \omega_\circ \text{ and } a_{xC_2} \equiv a_\circ \quad (2\text{-}22)$$

where $\omega_\circ$, $a_\circ$ are the proper quantities.

$$\omega a_1 = v_{1C_2} \equiv v_D \equiv v, \; \omega a_2 = v_{2C_2} \equiv v_B, \text{ and } \Gamma \equiv c \quad (2\text{-}23).$$

Owing to:

1- Unrecognizing of $\omega_1$ and formulate the system without it, and

2- regarding the ratio $\mu$ with the relative gear space; Minkowski space is obtained.

So VGM may lead to the relativistic formulation for a particle of a relative energy ($\hbar\omega$).
These relativistic formulations does not exhibit any feature of a gear system.

# 4. Conclusion

The relative consideration for the gear is equivalent to the observation process. The observable entity is particle of rest mass. Accordingly, there is no gear model can be observed. That system is a virtual gear model.